\documentstyle{cupconf} 
\newcommand{\ha}{H$\alpha$}

\newcommand{\hi}{H~{\sc i}}

\newcommand{\hei}{He~{\sc i}}

\newcommand{\nii}{[N~{\sc ii}]}

\newcommand{\Oi}{O~{\sc i}}
\newcommand{\feii}{Fe~{\sc ii}}
\newcommand{\mgii}{Mg~{\sc ii}}
\newcommand{\caii}{Ca~{\sc ii}}
\newcommand{\SiII}{Si~{\sc ii}}
\newcommand{\nai}{Na~{\sc i}}

\title[Circumstellar interaction in Type~II supernovae]{Supernova progenitor 
constraints from circumstellar interaction: Type~II}

\author[Cumming \& Lundqvist]%
{R\ls O\ls B\ls E\ls R\ls T\ns J.\ns C\ls U\ls M\ls M\ls I\ls N\ls
G\ns
\and 
\ns P\ls E\ls T\ls E\ls R\ns L\ls U\ls N\ls D\ls Q\ls V\ls I\ls S\ls T}
\affiliation{Stockholm Observatory, S-133~36 Saltsj\"obaden, Sweden 
(e-mail robert, peter@astro.su.se)\\[\affilskip]}

\large
\begin{document}

\maketitle

\begin{abstract}

All types of supernovae (SNe), except Type Ia (see Lundqvist \&
Cumming, this volume) have been observed to interact with their
immediate circumstellar medium (CSM).  This interaction can reveal
their progenitor's histories, and in a broader sense constrain our
ideas about the evolution of massive stars.  Progress in this
direction has concentrated itself in three areas.  First, radio
emission has been detected from a variety of core-collapse SNe, and
modelling according to the simple model of Chevalier (1982) has given
estimates of progenitor mass loss rates.  For some of these SNe,
especially SN 1993J, complementary information comes from X-ray
observations.  Second, multiwavelength observations of SN 1987A in the
LMC have shown that the progenitor star passed through a phase as a
red supergiant with a highly asymmetric and nitrogen-enhanced wind.
Third, optical observations of some core-collapse SNe show that their
emission to a large extent comes from circumstellar interaction,
indicating that they are surrounded by a dense CSM.

As examples of the latest progress in these areas, we present some new
results on two SNe: the famous 1987A, and the less well-known but
rather intriguing 1994W.  In the process we introduce Type IIn
supernovae (SN IIn; Schlegel 1990) as an important new probe of
massive star evolution.

\end{abstract}

\firstsection 

\section{SN 1987A: a 3D map of the triple-ring nebula}

In \cite{CLMSA97} we present and model high-resolution
($R\sim$50\,000) spectra of SN 1987A's three circumstellar rings,
taken with the \'echelle spectrograph UCLES at the Anglo-Australian
Telescope between 1990 and 1993.
We measure the expansion velocity for the inner ring to be
10.3$\pm$0.4 km~s$^{-1}$; here we present the most detailed
measurements yet of the kinematics of the two outer rings (Figure
\ref{f-3d}).
  \begin{figure} \centering \vspace{9.5cm}
  \includegraphics{cumming_lundqvist_fig1.ps} \caption[3D map of the
  3-ring nebula]{Four views of the three-dimensional structure of the
  triple-ring nebula, assuming homologous expansion.  The morphology
  of the \nii\ nebula hardly changed between 1036-1790d
  (Wang \& Wampler 1992) and the $\sim$2550-d
  {\sl HST} imaging (Burrows \etal\ 1995).  Using these images as a
  guide, we measured the velocities in different parts of the nebula
  from our position-velocity plots for days 1344 and 1397 together
  with the day 1402 observations of \cite{CH91}.  Then, assuming
  homologous expansion, we mapped each measured point to a position
  along the line of sight according to its velocity.  Hydrodynamic
  models presented so far for the SN 1987A CSM have velocity fields
  not too different from this (e.g., Blondin \& Lundqvist 1993).  
  We followed the same procedure for the short-lived \hi\ and \hei\
  blobs (Cumming \& Meikle
  1993 and references therein; open circles).  Broken lines join points
  connected to each other in {\sl HST} imaging.  }
\label{f-3d} \end{figure}

The outer rings seem to be parallel to and equidistant from the inner
ring.  Each lies 0.4$\pm$0.1($D_{\rm LMC}$/50~kpc) pc from the inner
ring plane.  Comparing with the continuum echo mapping of the dusty
shell between the rings (Crotts, Kunkel \& Heathcote 1995), we see that the
separation of the rings is close to the distance to the polar edges of
the shell, 0.6($D_{\rm LMC}$/50~kpc) pc.  Given the errors, the rings
coincide with the outermost edges of the dusty shell.  Our results
strengthen the picture of the nebula as an open-ended,
hourglass-shaped shell and provide an important constraint for
hydrodynamic and stellar evolution models (e.g., Arnett, this volume).

\section{SN 1994W: a type IIn supernova}

We present observations from the first 201 days of the remarkable
narrow-line Type II supernova 1994W in NGC 4041, taken at the {\sl
WHT, INT, JKT} and {\sl NOT} on La Palma.  A full analysis will appear
in \cite{CLSP97}.  In Figure \ref{f-lightcurve} we show the available
photometry for SN 1994W.  Figure \ref{f-spectrum} shows the spectral
development.
  \begin{figure} \centering \vspace{8cm}
  \includegraphics{cumming_lundqvist_fig2.ps} \caption[Light curve of SN 1994W]{Light
  curve of SN 1994W.  The inverted triangles are visual magnitude
  estimates by L. Szentasko which we have corrected to $V$ by
  comparing a wide selection of his observations with contemporary
  photoelectric $V$ measurements.   The drop in the light curve after day 100 
  is the fastest ever seen in an SN, and is nicely bracketed by 
  photometric points from \cite{T95} and our day 124 $V$ point.
}  \label{f-lightcurve} \end{figure}
  \begin{figure}
  \centering
  \vspace{8cm}
  \includegraphics{cumming_lundqvist_fig3.ps}
  \caption[Spectral evolution of SN 1994W]{Spectral evolution of SN
  1994W.  On days 31 and 57 the spectrum was dominated by narrow
  ($\la 10^3$ \kms) \hi\ lines with P-Cygni profiles and broad wings 
  ($\sim\pm10^4$ \kms), which were accompanied by low-ionisation,
  mainly narrow P-Cygni lines --- we identify \hei\ (with
  triangular profiles), \Oi,
  \mgii\, \SiII\ and \feii, but no forbidden lines.  The
  light curve drop coincided with a dramatic change in the
  spectrum.  On day 120, just after the fading, \ha\ had narrowed and
  was present only in emission, accompanied by faint \nai\ and \caii.
  By day 201, only \ha\ was left.}
  \label{f-spectrum} \end{figure}

The presence of the narrow P-Cygni lines seems quite consistent with
the clumpy wind model presented by \cite{CD94} and Chugai, Danziger \&
Della Valle (1995) for SNe 1988Z and 1978K.  In this scenario, the
wind consists of dense clumps embedded in a rarefied interclump wind.
The broad component in \hi\ is most easily explained using the shock
model of \cite{CF85}, in which the line wing emission comes from the
cool, shocked ejecta close to the contact discontinuity between the
shocked ejecta and the shocked circumstellar wind.  In the clumps, the
shock is slowed down, and high column density in the clouds can
account for the observed narrow P-Cygni profiles.  As long as the SN
is covered by the clouds, we cannot see the broad \ha\ component
normally seen in Type II SNe.  The absence of normal ejecta lines in
SN 1994W suggests both that the clumpy wind prevents us from seeing
them, and that a substantial proportion of the SN's energy is
reprocessed into circumstellar emission.  We suggest that the drop of
the light curve around day 120 marked the point at which the outgoing
shock had passed most of the clumps.  The simultaneous fading of the
lines and the continuum indicates that {\em both} had a circumstellar
origin.

SN 1994W's progenitor probably lost mass in a clumpy wind shortly
before it exploded, a picture similar to the one suggested for SN
1984E (e.g., Gaskell \& Keel 1988).  However, SN 1984E showed normal
ejecta lines both while its narrow lines were present and after they
had disappeared.  The resulting circumstellar shell round SN 1994W
had, in contrast to 1984E, a clump filling factor high enough to
conceal the SN ejecta entirely, and to reprocess most of its
radiation.

\subsection{The diversity of Type IIn supernovae: a picture emerges}

We suggest that the clumpy-wind model can be extended to account for
the general features of {\em all} SN IIn, without abandoning spherical
symmetry.  It is the clumps that produce the narrow lines which
distinguish these SNe as a class.  The broad lines come from the SN
ejecta's interaction with the interclump wind material.  The
prominence of the narrow lines relative to broad lines (either from
the SN photosphere, as in SN 1984E, or from the interaction region, as
in SNe 1988Z and 1994W) depends on the covering and filling factors of
the clumpy wind, as well as the density ratio of the clump to
interclump gas.  Unlike SNe 1984E and 1994W (and probably SN 1987B;
Schlegel \etal\ 1996), the progenitors of the
slow SN IIn like SNe 1988Z, 1986J and 1978K (Leibundgut 1994, and
references therein) had long-lived clumpy winds which still haven't
been passed by the expanding shock.

In SN IIn, the flux evolution of the narrow lines measures the
duration of clumpy mass loss ($\sim$500 yr for 1994W, for a 10-\kms\
wind; much longer for 1988Z), while their widths measure the density
contrast between the clumps and the interclump wind, if we know the
original shock speed.  That can be estimated from the width and
evolution of broad lines at early times.  From the early evolution of
some SN IIn, like SN 1994W, we can probe the evolution of a massive
star just prior to explosion.  SN IIn are a whole class of supernovae
that have the potential to provide us with just as detailed tests of
stellar evolution as SN 1987A has.

\noindent {\bf Acknowledgements.}  We are happy to acknowledge
stimulating collaboration with Peter Meikle and Jason Spyromilio (SN
1987A) and Jesper Sollerman and Miguel P\'erez Torres (SN 1994W).  We
also thank Laszlo Szentasko for visual estimates, and Mike Breare,
Ren\'e Rutten and Phil Rudd for service spectra and photometry of SN
1994W.

{} 

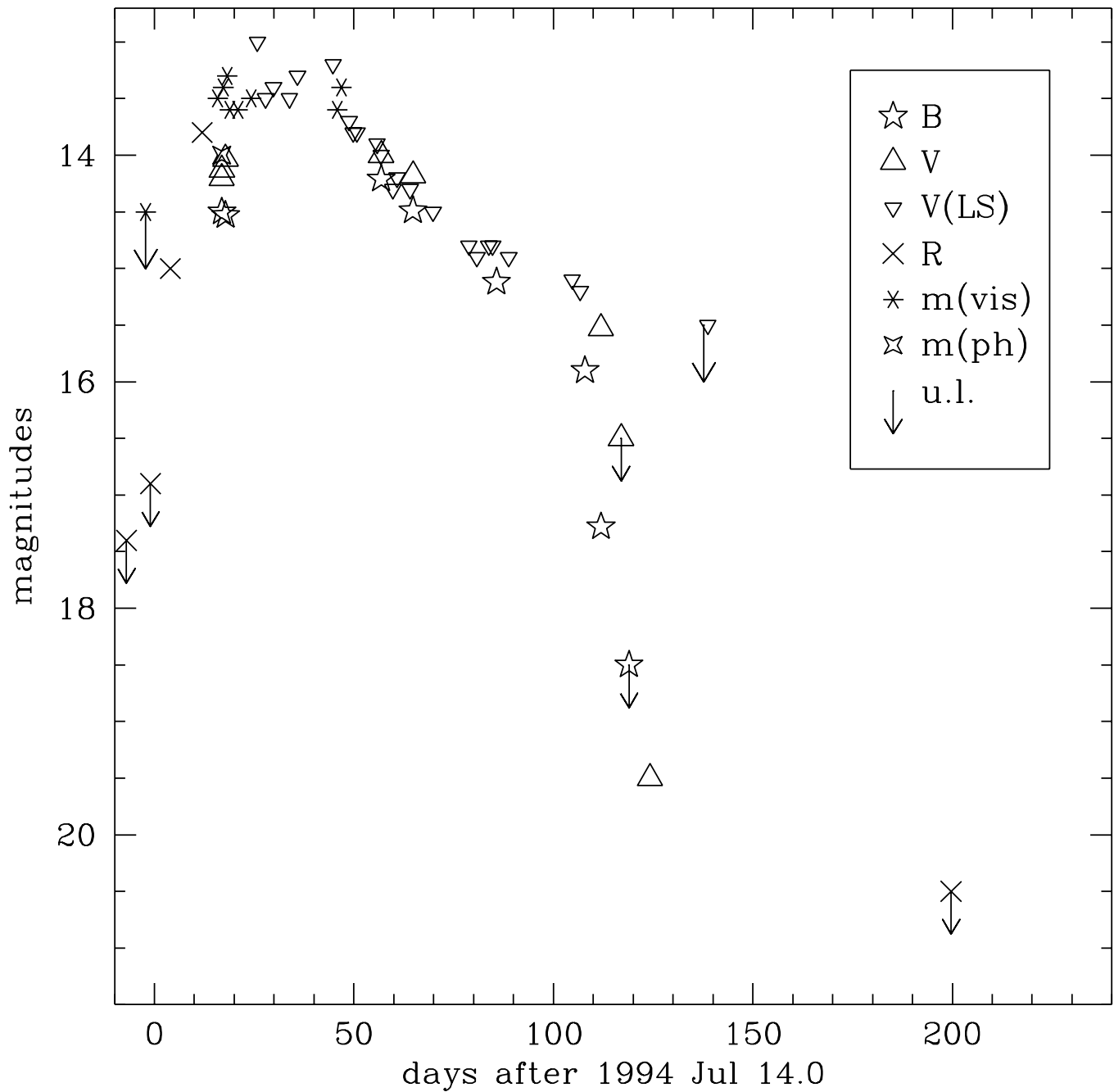
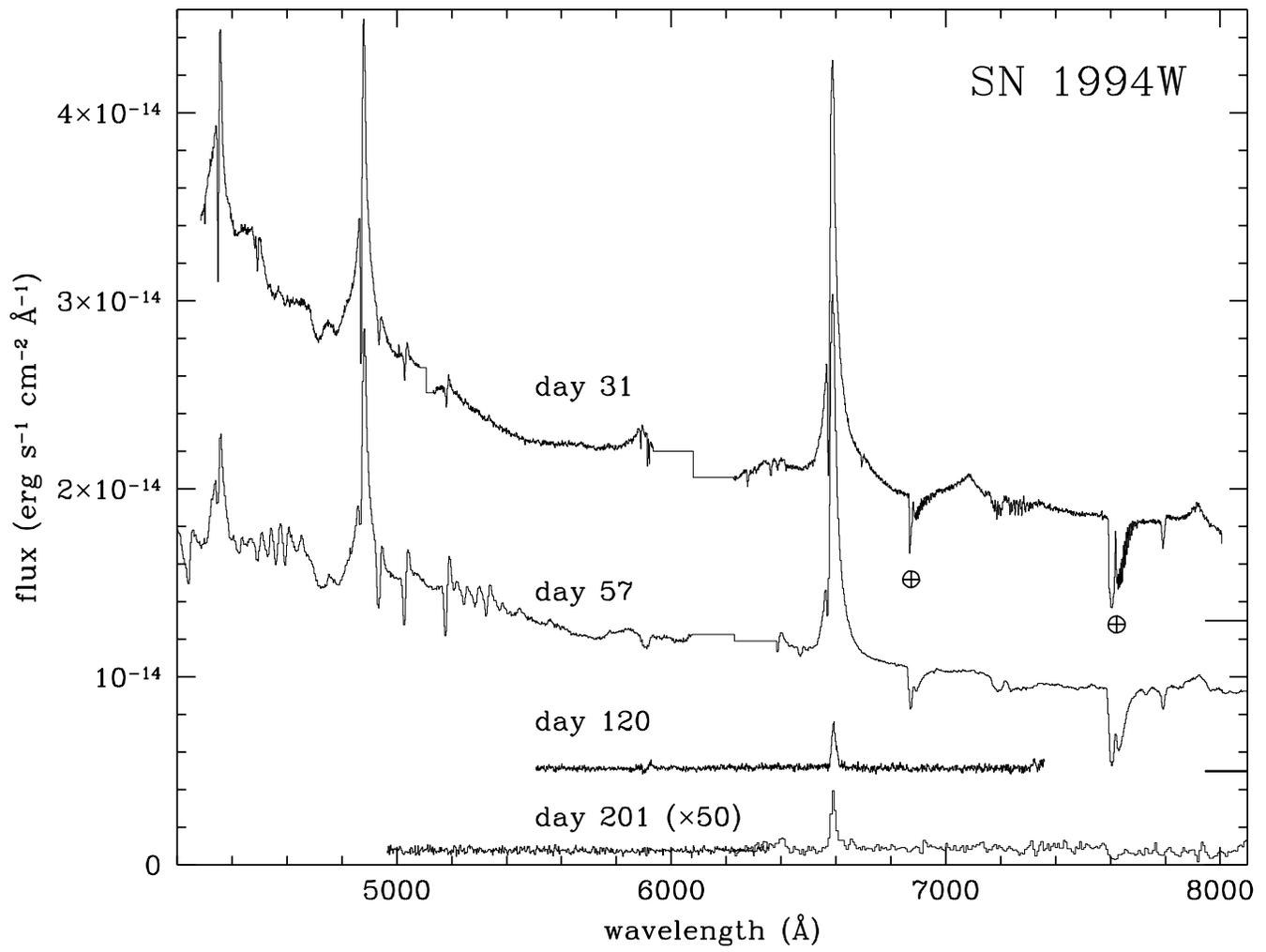
\end{document}